\documentclass[prl,aps,twocolumn]{revtex4}

\usepackage{graphics}
\begin{document}

\title{\Large Quantum imaging of nonlocal spatial correlations \\ induced
by orbital angular momentum}

\author{Adam R. Altman, Kahraman G. K\"{o}pr\"{u}l\"{u}, Eric Corndorf, Prem
Kumar, and Geraldo A. Barbosa}

\affiliation{Center for Photonic Communication and Computing, ECE Department \\ Northwestern University, Evanston, IL
60208-3118 }

\email{Email: g-barbosa@northwestern.edu}

\begin{abstract}
Through scanned coincidence counting, we probe the quantum image produced by parametric down conversion with a pump beam
carrying orbital angular momentum. Nonlocal spatial correlations are manifested through splitting of the coincidence spot
into two.
\end{abstract}

\maketitle

\noindent It is well known that an optical-parametric down converter---an optical-parametric amplifier with no signal and
idler inputs---produces twin photons that are entangled in energy and momentum \cite{pqitext}. The down-converted twin
photons at the output of the nonlinear crystal form cones and the angle between the wavevectors of the down-converted
photons and the pump is determined by phase matching conditions. The signal and idler cones overlap for a degenerate
($\lambda_s=\lambda_i$) type-I $\chi^{(2)}$ crystal and if the pump is a plane wave, or a Gaussian beam, the wavevectors of
the twin photons and the pump lie in the same plane. Simultaneous pair-wise creation of the signal and idler photons leads
to nonlocal correlations, which are manifested in coincidence detection of the twin photons. Keeping one detector fixed and
the other scanning yields a single-spot pattern whose spatial extent depends on the pump-waist size and the phase-matching
characteristics of the crystal. However, it has been shown theoretically \cite{barbosa-hugo,barbosa} that if the crystal is
pumped by a beam carrying orbital angular momentum (for example, a Laguerre Gaussian beam with $l \neq 0$), then the twin
photons propagate off-plane with respect to the pump beam's central wave vector. The off-plane deviation is dependent on the
value of $l$ and, therefore, provides a measure of the pump's orbital angular momentum. Because of this off-plane angle, a
nonlocal spatial pattern is predicted in coincidence detection of the twin photons. With one detector fixed and the other
scanning, the usual single-spot coincidence pattern is predicted to split into two spots whose separation depends on the
value of $l$. This nonlocal spatial splitting phenomenon is a new type of quantum image that is produced by twin photons
entangled in orbital angular momentum. In this paper, we demonstrate, for the first time to our knowledge, an experimental
demonstration of such a predicted quantum image.

\begin{figure}
\centerline{\scalebox{0.35}{\includegraphics{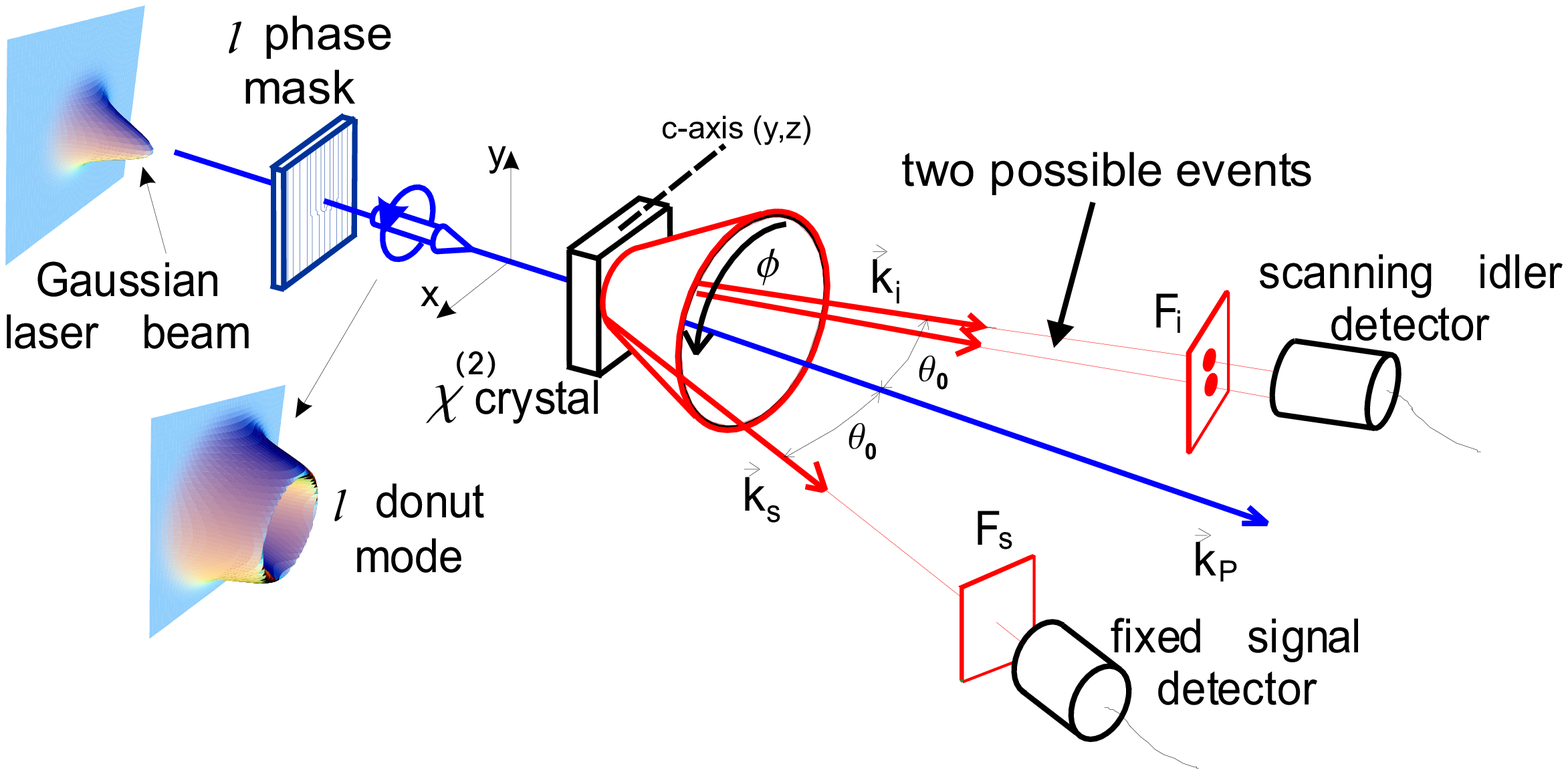}}} \caption{A sketch of the experimental setup. A laser beam
with $\lambda = 351$\,nm is holographically prepared in $l=4$ state of orbital angular momentum and focused on a
$\chi^{(2)}$ nonlinear crystal cut for type-I phase matching. One member of the down-converted photon pairs (signal photons)
is detected by a fixed photon-counting detector while the conjugate photons (idler photons) are collected by a scanning
photon-counting detector. $F_s$ and $F_i$ are optical interference filters centered at $702$\,nm with $10$\,nm bandwidth. }
\label{Lexperiment}
\end{figure}

\begin{figure}
\centerline{\scalebox{0.35}{\includegraphics{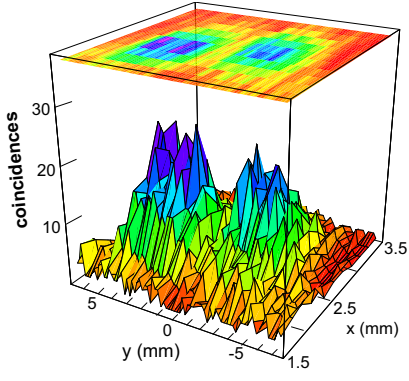}}} \caption{Two coincidence spots are observed as the idler
detector is scanned. The fixed detector is placed at $|x_s| = 3.7$\,cm and $y_s = 0$\,cm from the pump-beam axis. Pumping
with a TEM$_{00}$ Gaussian mode produces only a single coincidence spot at $y = 0$.} \label{coincidences}
\end{figure}

Figure~\ref{Lexperiment} sketches the down-conversion geometry we utilize and the resulting two-spot pattern is shown in
Fig. \ref{coincidences}. We use the 351.1\,nm output of a CW argon laser as the pump beam, which is passed through a
holographic phase mask to produce a beam with $l = 4$. The resulting Laguerre-Gaussian beam with a donut shape is focused
into a 2\,mm-long BBO crystal cut for type-I phase matching ($\theta = 35.2^\circ$, $\phi = 90^\circ$) by use of a 17.5\,cm
focal-length convergent lens. The pump beam after the crystal is blocked by a filter and a beam stop. The down-converted
photons are collected by two photon-counting modules (PCMs) having a detector area of diameter 175\,$\mu$m. Optical
interference filters centered at 702\,nm having 10\,nm bandwidth are used in front of both PCMs. No lenses or apertures are
used between the crystal and the detectors, allowing us to obtain a substantially higher-resolution image at the cost of
lower counting rates. The scannable PCM is mounted on a computer-controlled, stepper-motor driven, translation stage with a
minimum step size of 10\,$\mu$m. The single and coincident counts are recorded with a pair of two-channel photon counters.
One counter is operated in a gated mode and registers the photon clicks from the fixed detector for a 5-ns duration (the
smallest gate-time possible with our photon counter) when triggered by the movable detector in response to a
photon-detection event. This allows us to measure the coincidence counts between the detectors at two points with minimum
number of accidental counts. The second counter simultaneously registers the number of triggers from the movable detector,
yielding the intensity pattern of the parametric down-conversion ring in the detection plane.

\begin{figure}
\centerline{\scalebox{0.4}{\includegraphics{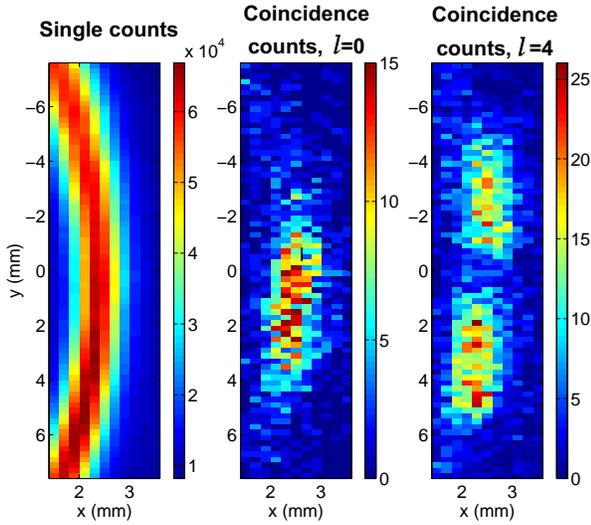}}} \caption{Left---Single counts for idler within a sector of the
down conversion ring for a pump beam with $l=0$, pump power $200$mW over 10s. Middle---Coincidences for pumping with a
Gaussian beam ($l=0$) and pump power $200$mW. Right---Split-spot coincidences for a pump beam with $l=4$ and pump power
400mW over 10s.}\label{sl0cl0cl4}
\end{figure}

\begin{figure}
\centerline{\scalebox{0.35}{\includegraphics{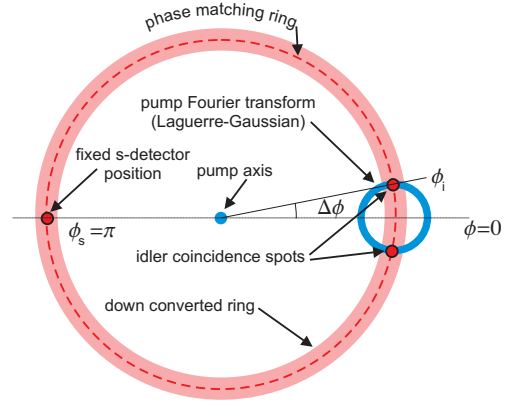}}} \caption{A detection-plane representation of the pump's
donut-mode Fourier transform intersecting with the down-conversion ring, which gives rise to the observed split-spot pattern
shown in Fig.~\ref{sl0cl0cl4}(c). We define $\Delta \phi$ as the aperture of the split-spot pattern.} \label{overlap}
\end{figure}

Figure~\ref{sl0cl0cl4}(a) shows a portion of the ring of down-converted photons mapped by the single-count output from the
scanning idler PCM. Figures \ref{sl0cl0cl4}(b) and (c), respectively, show the spots mapped by coincident counts when the
crystal is pumped by a Gaussian-intensity-profile pump ($l=0$) and by a pump beam carrying orbital angular momentum ($l=4$).
Clearly, in the latter case the coincidence spot is split into two. Our theoretical analysis shows that the image produced
by coincidence counts is the Fourier transform of the pump field inside the nonlinear crystal. However, we are able to
observe only a portion of this image since the parametric down converter has a spatial bandwidth owing to the phase-matching
requirement. Since the Fourier transform of a Laguerre-Gaussian (donut profile) is also a Laguerre-Gaussian, what we see in
the detection plane is that part of the donut which falls within the phase-matching bandwidth. To better visualize this,
consider the quantum state of the down-converted photons with orbital angular momentum (Eq. (21) in Ref.
\cite{barbosa-hugo}) for two wavevectors ${\bf k}_s$ and ${\bf k}_i$:
\begin{eqnarray}
\label{wfunction} |\psi_{lp}(t)\rangle=\sum_{n=-l}^{l}F_{lp}^{(n)}(\Delta{\bf k})\hat{a}_{j_z=n}^\dagger({\bf
k}_s)\hat{a}_{j_z=l-n}^\dagger({\bf k}_i)|0\rangle \,,
\end{eqnarray}
where $\Delta {\bf k}={\bf k}_s+{\bf k}_i-{\bf k}_P$ is the wavevector off-set, $F_{lp}^{(n)}(\Delta{\bf k})\propto \int
d^3r \, \psi_{lp}({\bf r})e^{-i \Delta {\bf k}\cdot {\bf r}}$, and $\psi_{lp}$ is the Laguerre-Gaussian pump (donut) mode.
For a crystal thickness much smaller than the Rayleigh range, $l_c\ll z_R$, $F_{lp}^{(n)}(\Delta{\bf k})$ can be written as
\begin{eqnarray}
F_{lp}^{(n)}&\propto &\int dz \, e^{-i z \Delta {\bf k}_z} \nonumber \\& &\times \int dx dy \, \psi_{lp}(x,y)\, e^{-i(x
\Delta {\bf k}_x+y \Delta {\bf k}_y )} ,
\end{eqnarray}
wherein first integral represents the phase-matching constraint, giving the polar angle for the down-conversion cone, and
the second is a 2-d Fourier transform of the Laguerre-Gaussian pump mode. This Fourier transform is also a Laguerre-Gaussian
mode and, therefore, the product of the two integrals is a two-spot pattern resulting from the overlap of the
down-conversion cone with the donut mode, as schematically shown in Fig.~\ref{overlap}. This two-spot pattern is what is
observed upon placing the detectors in the far field as shown in Fig.~\ref{sl0cl0cl4}(c).

\begin{figure}
\centerline{\scalebox{0.4}{\includegraphics{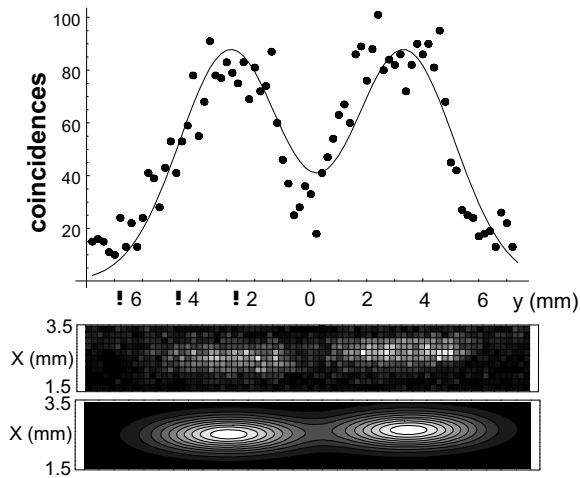}}} \caption{Histogram of the coincidence pattern projected along
the $y$ direction (top) and a two-dimensional Gaussian function (bottom) fitted to the experimental data (middle).}
\label{yprofile}
\end{figure}

As detailed in Ref.~\cite{barbosa-hugo}, phase matching gives the following longitudinal and transversal conditions:
\begin{eqnarray}
-k_P+(k_s \cos \theta_s +k_i \cos \theta_i)=\frac{2 \pi}{l_c} \nonumber \:,\hspace{3cm} \\ (k_s \sin \theta_s)^2+(k_i \sin
\theta_i)^2+2 k_s \sin \theta_s k_i \sin \theta_i \cos(\phi_s-\phi_i)\nonumber\\=\frac{k_P l}{z_R} \:. \nonumber
\end{eqnarray}
These equations for the degenerate case ($\omega_s = \omega_i$) provide the scattering polar angle
\begin{eqnarray}
\theta=\theta_s=\theta_i=\arccos\left(\frac{k_P+(2 \pi/l_c)}{2 k_s} \right)\:\:,
\end{eqnarray}
and the split-spot aperture $\Delta \phi \equiv |\phi_s-\phi_i|$ given by
\begin{eqnarray}
\Delta \phi=\arccos\left( \frac{\mbox{cosec}^2 \theta \left[ k_P l-(2 k_s^2 z_R \sin^2\theta) \right]}{2 k_s^2 z_R} \right)
\:\:.
\end{eqnarray}
The radius of the donut beam, and hence the separation between the coincidence spots, increases with $l$. Therefore, one can
calculate the value of $l$ from the coincidence data, using the constant of motion $|(k_{sx}+k_{ix})+(k_{sy}+k_{iy})|^2=k_P
l/z_R$ (Eqs. (16) and (27 ) in Ref. \cite{barbosa-hugo}). For the data presented in Fig.~\ref{yprofile}, we obtain $\Delta
y_0=3.3$mm. That together with the experimental parameters $\theta_s=0.0698$, $n_s=1.6648$, $k_s=1.4897\times
10^5$cm$^{-1}$, $z_R=0.5$cm, $k_P=2.9719\times 10^5$cm$^{-1}$, and the measured down-conversion ring radius of $R=3.7$cm in
the detection plane, give through the constant of motion:
\begin{eqnarray}
l=\!\frac{z_R}{k_P}\left( k_s \sin\theta_0 \right)^2 \!\!\left[ \!\left(1-  \sqrt{1-\!\!\frac{\Delta y_0^2}{R^2} }
\right)^2\!\!+ \left( \frac{\Delta y_0}{R \cos \theta_0}\right)^2\! \right]\!\!\simeq 4 \nonumber.
\end{eqnarray}
Measurements done with an $l=2$ mask did not distinctly reveal the separated peaks owing to the noise level at the
achievable coincident-counting rates in our experiment. 

\begin{figure}
\centerline{\scalebox{0.7}{\includegraphics{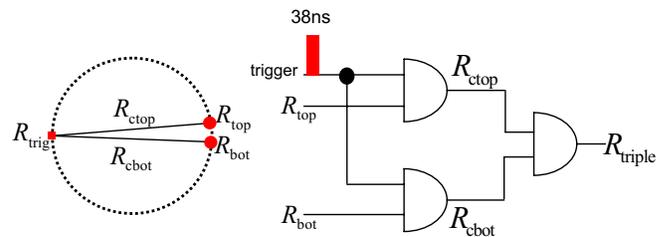}}} \caption{Schematics of the coincidence spots (left) and the
electronic circuitry used for triple coincidence measurements (right). $\Delta t=38$\,ns is the trigger pulse width. }
\label{triple}
\end{figure}

In order to determine whether or not this splitting is a quantum effect, we also measured the triple coincidences, i.e., we
looked for simultaneous coincidences between the fixed detector and the two parts of the two-spot pattern with use of an
additional PCM. See Fig.~\ref{triple}. According to the quantum theory \cite{barbosa-hugo}, there should not be any such
triple-coincidence counts. However, semiclassically, the two coincidence spots may be considered to be generated by two
independent parametric down-conversion processes, wherein there would be coincidence events occurring simultaneously in both
parts of the two-spot pattern. The rate of triple counts in such a case is related to the rate of coincidences through
\begin{equation}\label{tcrate}
R_{\rm triple} = \frac{R_{\rm ctop} R_{\rm cbot}}{R_{\rm trig}},
\end{equation}
where $R_{\rm ctop}$ ($R_{\rm cbot}$) is the coincidence-count rate between the fixed detector and the detector looking at
the top (bottom) spot, and $R_{\rm trig}$ is the single-photon count rate at the fixed detector with respect to which the
coincidences are measured. To measure the rate of triple-coincidence counts, we use the same experimental setup as in
Fig.~\ref{Lexperiment}, but this time---instead of using the scanning detector---we use two fiber-coupled GRIN lenses that
are placed on the coincidence spots and connected to fiber-coupled PCMs. The advantage of this setup is that the GRIN lenses
have a much larger area than the 175\,$\mu$m diameter detectors, allowing us to obtain substantially higher
coincidence-count rates to acquire statistically meaningful triple-coincidence data in a manageably short period of time. A
simple digital circuit composed of AND gates is used to measure $R_{\rm ctop}$, $R_{\rm cbot}$, and $R_{\rm triple}$.

\begin{table}[t]
\centering
\begin{tabular}{|l|r||l|r|}
\hline
  $R_{\rm trig}$ & 564/s & $R_{\rm triple}$ & 0.0076/s \\\hline
  $R_{\rm ctop}$ & 4.1/s & $R_{\rm top}$ & 21479/s \\\hline
  $R_{\rm cbot}$ & 2.7/s & $R_{\rm bot}$ & 22486/s \\ \hline
\end{tabular}
\caption{Count rates when GRIN lenses were used instead of the scanning detector.}\label{crates}
\end{table}

With the above triple-coincidence counting setup, we took data over a period of $10^4$\,s and recorded the various count
rates. Table~\ref{crates} shows the measured values. The table also gives $R_{\rm top}$ and $R_{\rm bot}$, which are the
single-photon count rates at the top and the bottom GRIN lenses, respectively. The triple-count rate listed in
Table~\ref{crates} not only contains the true triple-count events that are caused by two simultaneous coincidences in the
top and bottom spots, but also the accidental triple counts because of the finite-duration gate time over which we make the
measurements. The rate of such accidental triple counts is
\begin{equation}\label{acc}
R_{\rm acc}=(R_{\rm ctop} R_{\rm bot} + R_{\rm cbot} R_{\rm top}) \tau / 2,
\end{equation}
where $\tau$ is the gate time. Our circuit uses AND gates to find the overlap between 38\,ns pulses produced by the PCMs;
therefore, it is reasonable to assume a gate time of $2\times 38$\,ns in our calculations.

Using Eq.~(\ref{acc}) and the values in Table~\ref{crates}, $R_{\rm acc}$ is calculated to be 0.0057/s. If we subtract this
value from the measured value of the triple-count rate, we find the rate of true triple counts to be 0.00189/s.

In contrast, using the semiclassical picture, the triple-count rate calculated via Eq.~(\ref{tcrate}) would be 0.0196/s,
which is $\sim10$ times higher than the rate of true triple counts measured in our experiment. The difference between the
two values shows that a coincidence event happening at one spot of the two-spot pattern strongly prevents a simultaneous
coincidence event to take place at the other spot, proving that the spatially-split coincidence pattern is a quantum effect
that cannot be explained by semiclassical theories.

The above results show that by placing two separate detectors in the two regions of the split spot, we are {\em not}
performing two separate experiments at the same time and, therefore, classical multiplication rule for probabilities should
not apply. The coincidence events giving rise to the observed pattern can be represented by the following entangled quantum
state:
\begin{eqnarray}
|\psi_{\rm split}\rangle=|1_{{\bf k}_s}\rangle \frac{\left[ |1_{{\bf k}_i}\rangle_{\mbox{\small top}}|0_{{\bf
k}_i}\rangle_{\mbox{\small bot}}+|0_{{\bf k}_i}\rangle_{\mbox{\small top}} |1_{{\bf k}_i}\rangle_{\mbox{\small bot}}
\right]}{\sqrt{2}}\:.
\end{eqnarray}
Such a state cannot be modelled by a hidden-variable theory and should violate all generalized Bell's inequalities.

Note that in the measurements presented in this paper, no phase masks were used in front of the detectors. Different weights
should be assigned in case such masks are used. To compare these cases, the wave function in Eq.~(\ref{wfunction}) can be
written as
\begin{eqnarray}
|\psi_{lp}(t)\rangle=G \sum_{n=-l}^{l} {l \choose m} |j_z=n\rangle_s |j_z=l-n\rangle_i \:\:,
\end{eqnarray}
where $G$ is a proportionality constant. One obtains the following projected wave function for the idler photons, if no
phase mask is used in the signal beam:
\begin{eqnarray}
|\psi_{lp}(t)\rangle_i &=& \sum_m \,_s\langle m|\psi_{lp}(t)\rangle \nonumber \\ &=& G \sum_m {l \choose m}
|j_z=l-m\rangle_i.
\end{eqnarray}
In case a $q$ phase mask is used in the idler path, the probability amplitude becomes
\begin{eqnarray}
\,_i\langle q|\psi_{lp}(t)\rangle_i = G  {l \choose q}\:\:;
\end{eqnarray}
whereas if no mask is used, the following probability amplitude results:
\begin{eqnarray}
\sum_q\,_i\langle q|\psi_{lp}(t)\rangle_i = G\: 2^l\:\:.
\end{eqnarray}


In summary, we have demonstrated a spatially-split nonlocal coincidence pattern at the output of a parametric down converter
pumped by a beam carrying orbital angular momentum. We have also shown that our experimental setup uses a quantum imaging
method wherein the image is stored on the pump beam itself. Although we have not looked for phase signatures in this work,
such characteristics of beams carrying orbital angular momentum can be explored as well. Furthermore, a donut-shaped pump
beam that carries orbital angular momentum is used in the present experiment. Various other kinds of pump beams can be used
to generate coincidence patterns with more complex spatial features for quantum-cryptography applications. Lastly, this
experiment also provides an independent verification of the orbital angular momentum entanglement reported in \cite{mair}
and predicted in \cite{hugo-barbosa}.

This research was supported in part by the U.S. National Science Foundation under grant PHY-0219382.

\end{document}